\documentclass[apsrev4,pra,superscriptaddress,twocolumn]{revtex4-1}  
\usepackage{graphicx}
\usepackage{amssymb,calc}
\begin{document}
\title{Far-off-Bragg reconstruction of volume holographic gratings: a comparison of experiment and theories.}
\author{Matej Prijatelj}
\affiliation{Jo\v{z}ef Stefan Institute, Jamova 39, SI 1001 Ljubljana, Slovenia}
\author{J\"urgen Klepp}
\affiliation{University of Vienna, Faculty of Physics, Boltzmanngasse 5, A-1090 Wien, Austria}
\author{Yasuo Tomita}
\affiliation{Department of Engineering Science, University of Electro-Communications, 1-5-1 Chofugaoka, Chofu, Tokyo 182, Japan}
\author{Martin Fally}
\email{martin.fally@univie.ac.at}
\affiliation{University of Vienna, Faculty of Physics, Boltzmanngasse 5, A-1090 Wien, Austria}
\date{\today,\jobname}

\begin{abstract} 
We performed light optical diffraction experiments on a nanoparticle-polymer volume holographic grating in an angular range including also far-off-Bragg replay. A comparison of three diffraction theories - on the same level of complexity - with our experimental results shows that the dynamical theory of diffraction and the first-order two-wave coupling theory using the beta-value method fit the data very well. In contrast, the prevalent two-wave coupling theory using the K-vector closure method yields a poor fit with an order of magnitude worse mean squared error. These findings must be considered for accurate determination of coupling strength and grating thickness. 
\end{abstract}
\maketitle
\section{Introduction}
It seems that after about eighty years on theories of diffraction from periodic structures everything should have been said and done. This is even more true since {\em Moharam} and {\em Gaylord} in a series of papers published a rigorous theory of diffraction (rigorous coupled wave analysis, RCWA), which covers nearly the entire scope of cases that have ever been relevant \cite{Moharam-josa81,Gaylord-apb82,Moharam-josa82,Moharam-josa82a,Moharam-josa83,Moharam-josa83.2,Gaylord-pieee85,Moharam-josaa86,Moharam-josaa95,Moharam-josaa95a}. However, for practical purposes approximate theories are still around, which are used instead. While the optics community has strongly opted for the coupled-wave approach of {\em Kogelnik} (K-vector closure method, KVCM) \cite{Kogelnik-Bell69} and much less for the one introduced by {\em Uchida} (Beta-value method, BVM) \cite{Uchida-josa73}, the neutron and X-ray communities are using solely the dynamical theory of diffraction (DDT) \cite{Rauch-78,
Ewald-adp17.1,Ewald-adp17.2,Ewald-adp16.1,Ewald-adp16.2,Zachariasen-45,Batterman-rmp64} or {\em Darwin's} early variant \cite{Darwin-PhilMag14a,Darwin-PhilMag14b}. Despite the fact that the discussion on diffraction theories for volume gratings might sound outworn, we will show that it is possible to experimentally single out a set of analytic theories which is significantly superior to others.

We identified three problems in the theories when treating the off-Bragg regime: The first is, that the KVCM misses to correctly account for energy conservation, the second is that KVCM and BVM only treat the half-space case, i.e., a single boundary. Thus boundary conditions for a parallel slab are not properly included and yield an ambiguity. Finally, with the DDT, which in principle yields the exact solution to the wave equation, typically an approximation of the dispersion surface for small off-Bragg conditions is performed in literature \cite{Russell-prep81,Batterman-rmp64,Rauch-78}, which leads to hyperbolic dispersion surfaces and is called hyperbolic approximation in what follows. It is stated by {\em Syms and Solymar} that the differences between the amplitudes derived from each of the theories are small \cite{Syms-ao83}. However, for geometrically thin gratings with large coupling coefficients as for our samples deviations in the far-off-Bragg regime are observable.

The result of this work is that the first order two-wave coupling theory employing the BVM as well as the DDT \cite{Russell-prep81} fit the experimental data extremely well, provided that we refrain from using the hyperbolic approximation for the latter. The KVCM fits the experimental data much worse. 

The paper is organized as follows: we start with a derivation of the relevant equations
for the two-wave coupling theories KVCM and BVM as well as for the modal DDT without the hyperbolic approximation. Then the experiment and the corresponding results are shown together with fits to each of the theories. We also compare the theories using proper approximations, discuss the implications and end with a conclusion.
\section{Diffraction from a sinusoidal volume grating}\label{sec:Theory}
In what follows we summarize the different approaches to solve the wave equation for sinusoidal transmission volume 
gratings and give the results for the diffraction amplitudes with a particular emphasis on the far-off-Bragg regime, which is subject to our experiments.

\begin{figure}
\centering
 \includegraphics[width=\columnwidth/4+\columnwidth/2]{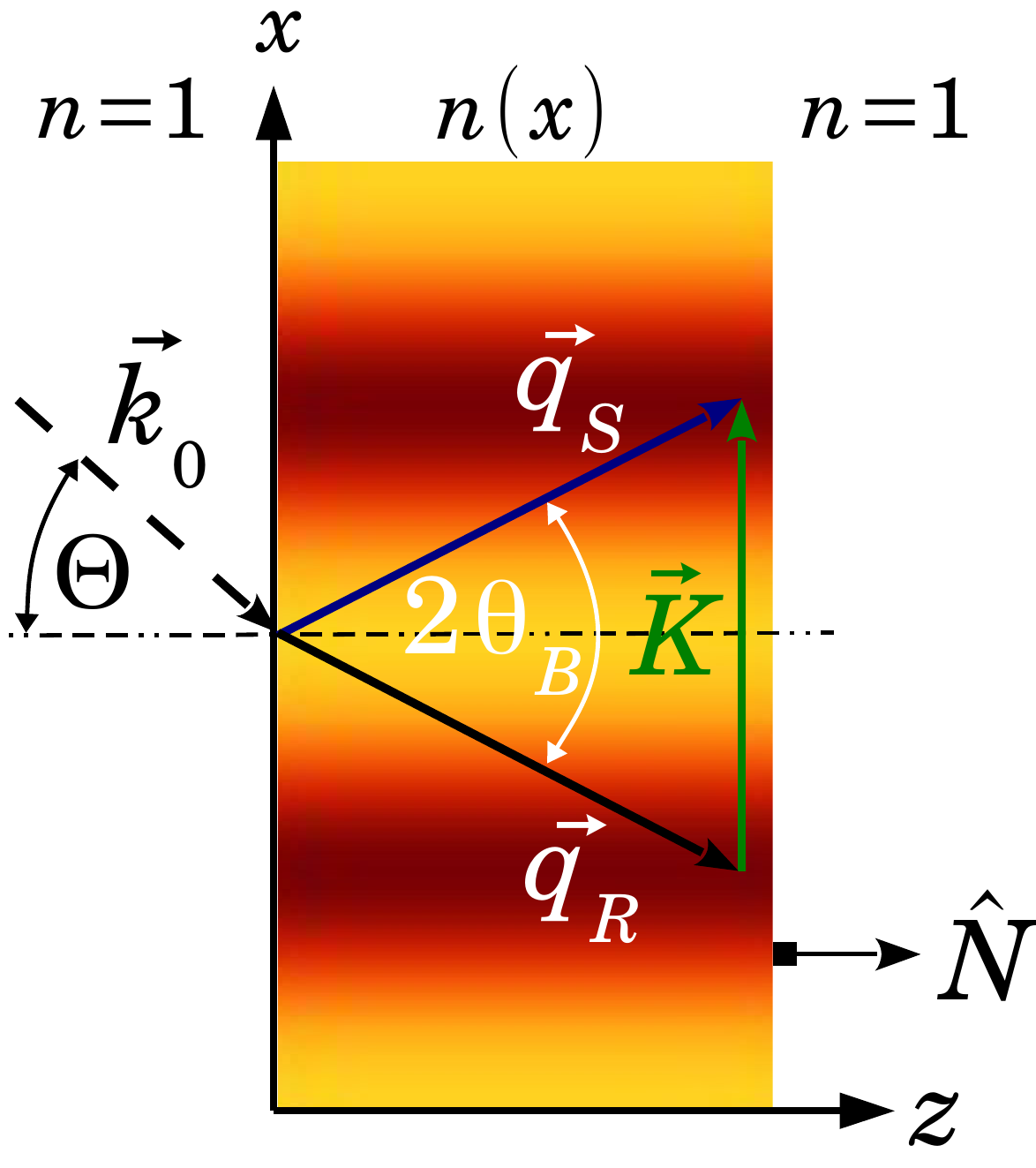}
 \caption{(Color online) Illustration of the geometry and the boundary condition. $\vec{k}_0,\vec{q}_R, \vec{q}_S$ are the wavevectors of the incident wave in the input region, the forward diffracted wave, and the diffracted wave in the grating region. $\vec{K}, \hat{N}, \Theta, \theta_B$ denote the grating vector, the sample surface normal unit vector, the angle of incidence (external) and the Bragg angle (in the medium).
 \label{fig:1}}
\end{figure}

We start with the simplest set of assumptions. Space is divided into three distinct regions: the input region (free space), the grating region (periodic material), and the output region (free space). A sketch of the geometry is provided in Fig. \ref{fig:1}.
The grating be
lossless,
one dimensional (modulated only along the $x$ direction), 
sinusoidal,
isotropic,
and of the pure phase-type.
Thus it can be described by
\begin{equation}\label{eq:grating}
n(x)=n_0+n_1 \cos{(K x)},
 \end{equation}
where $n_0$ is the mean refractive index of the material under investigation, $n_1$ the refractive-index modulation and $K$ the spatial frequency of the grating. The scalar wave equation (Helmholtz equation), for the polarization state perpendicular to the plane of incidence ($s$ or H mode polarization) in such a medium is
\begin{equation}\label{eq:waveequation}
 \left[\vec\nabla^2 +(n(x) k_0)^2\right]E(x,z)=0,
\end{equation}
where $k_0=2\pi/\lambda$ with $\lambda$ the free space wavelength of light and $E(x,z)$ the normalized electric field amplitude. It was shown that an exact solution of the wave equation can be obtained by solving an infinite number of coupled linear differential equations of first order under appropriate boundary conditions \cite{Moharam-josa81} (RCWA). In the rigorous treatment of the problem the coupled wave analysis and the modal approach are completely equivalent, i.e., lead to identical results \cite{Gaylord-apb82}. For thick volume holograms discussed here only two diffracted beams of considerable field amplitudes exist at the same time, e.g., the first and zeroth diffraction order \cite{Gaylord-ao81}. In addition, the amplitudes strongly depend on how well the Bragg condition, $2\beta\sin{\theta_B}=K$, is fulfilled. Here $\theta_B$ is the Bragg angle and $\beta=k_0 n_0$ the propagation constant in the material. A commonly defined off-Bragg parameter \cite{Kogelnik-Bell69}
 is
\begin{equation}\label{eq:OffBraggparameter}
 \vartheta=K\left(\sin{\theta}-\frac{K}{2\beta}\right),
\end{equation}
where $\theta$ is the angle of incidence.
Our interest is to compare three common diffraction theories with experimental data obtained for a wide range of angles, particularly in the far-off-Bragg regime.
In contrast, usually only the very vicinity of the Bragg peak is considered and a linearization of $\vartheta$ is performed (see also appendix \ref{app:2}), a simplification which is not applicable in our case. In what follows we give the relevant equations for each of the three theories that cover also the far-off-Bragg regime.
\subsection{First-order two-wave coupling: Kogelnik's approach (KVCM) \label{sec:Kogelnik}}
\begin{figure}
 \includegraphics[width=\columnwidth]{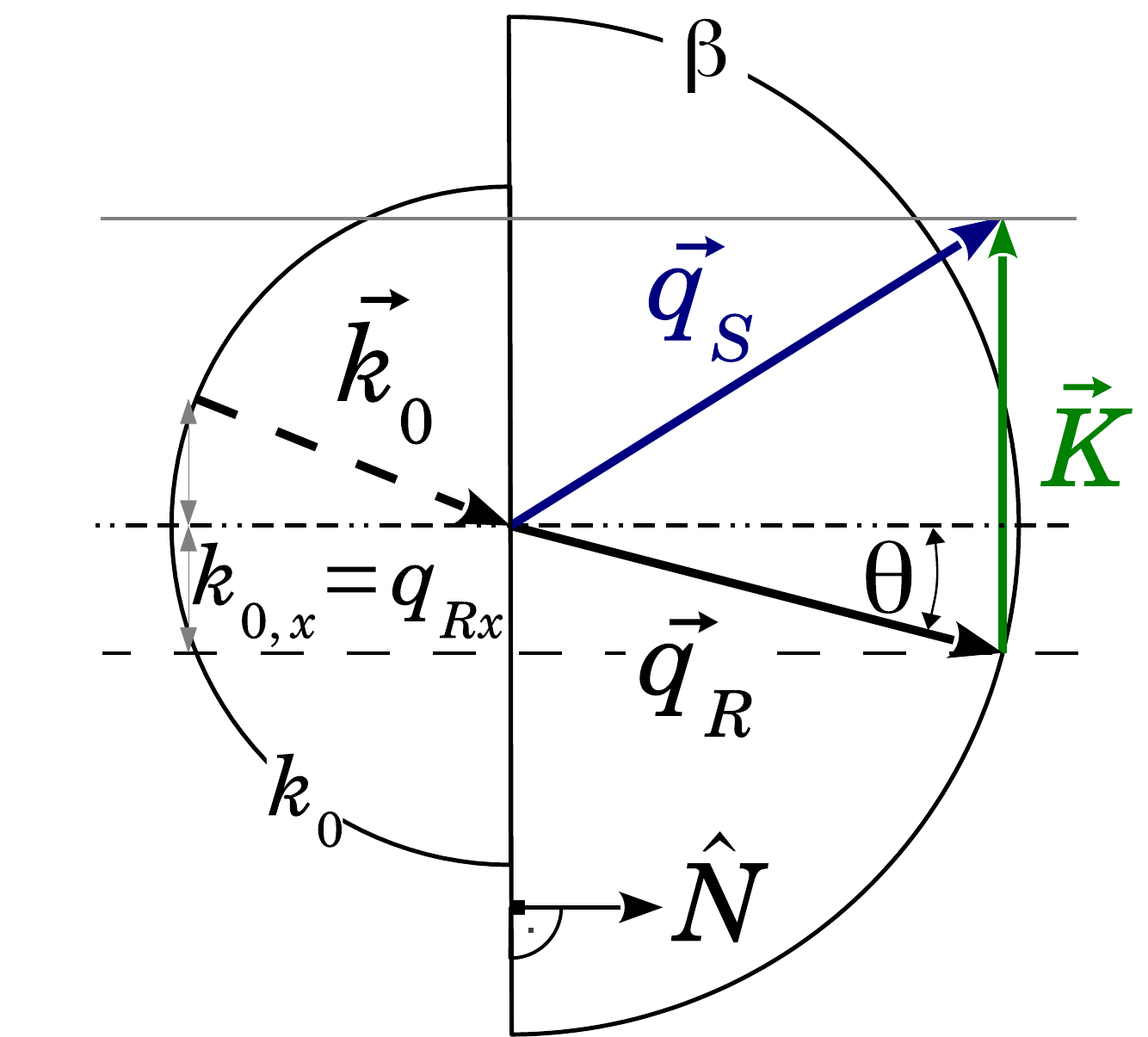}
 \caption{(Color online) Wavevector diagram for the KVCM (Kogelnik) in off-Bragg position with $\beta=|\vec{q}_R|\not=|\vec{q}_S|$ according to Ref. \cite{Kogelnik-Bell69}. $\hat{N}$ the surface normal unit vector.}
 \label{fig:wvKog}
\end{figure}

The basic idea of Kogelnik's coupled wave approach is to solve Eq. (\ref{eq:waveequation}) by the ansatz
\begin{equation}\label{eq:solansatz}
 E(x,z)=R(z)\exp{(\imath \vec{q}_R\cdot\vec{x})}+S(z)\exp{(\imath \vec{q}_S\cdot\vec{x})},
\end{equation}
i.e., the sum of two waves whose amplitudes vary when propagating through the sinusoidal refractive index pattern. By inserting Eq. (\ref{eq:solansatz}) into Eq. (\ref{eq:waveequation}) two coupled differential equations for the amplitudes $R(z), S(z)$ result that can be solved for appropriate boundary conditions. In most cases, second order derivatives can be neglected because the amplitudes are slowly varying functions as compared to the exponentials in Eq. (\ref{eq:solansatz}) (slowly varying envelope approximation). This simplifies the system of differential equations to first order on the expense that the boundary conditions are to some extent ambiguous (we end up with a half-space case instead of a slab, i.e., the output region is not considered). 
Coupling of the waves arises via the Bragg condition, which relates the wavevector $\vec{q}_S$ of the diffracted beam to the wavevector $\vec{q}_R$ of the forward diffracted - frequently called also 'transmitted' \footnote{For a clarification of subtle differences between the terms see e.g., Ref. \cite{Batterman-rmp64}} - beam. At this point {\em Kogelnik} introduces the relation
\begin{equation}\label{eq:kmatchKog}
 \vec{q_S}=\vec{q_R}\pm\vec{K}.
\end{equation}
While this is a correct choice from a mathematical point of view \cite{Syms-90}, it gives physically meaningful results only if the Bragg condition is fulfilled exactly. When going off-Bragg, Eq. (\ref{eq:kmatchKog}) predicts either a wavelength change (sort of 'inelastic' scattering) that does, in fact, not occur, or a refractive index that strongly depends on the off-Bragg parameter. Furthermore, the direction of the diffracted beam's wavevector is not correctly predicted for the off-Bragg case \cite{Fally-apb12}. A wavevector diagram for the off-Bragg case using Kogelnik's choice (KVCM) is shown in Fig. \ref{fig:wvKog}. 
The diffraction efficiency for the first order, the quantity measured in our experiment, is given by 
\begin{eqnarray}\label{eq:DEKogelnik}
\eta_{K}&=&SS^*=\left[\nu_K {\rm sinc}\sqrt{\nu_K^2+\xi_K^2}\right]^2\\
\nu_K&=&\kappa\frac{d}{c_R} \label{eq:nuKogelnik}\\
\xi_K&=&\vartheta\frac{d}{2 c_R} \label{eq:xiKogelnik}
\end{eqnarray}
with $\kappa=n_1\pi/\lambda$ the coupling coefficient, $d$ the hologram thickness, $c_R=\cos\theta$ and $^*$ denotes the complex conjugate.
\subsection{First-order two-wave coupling: Uchida's approach (BVM)\label{sec:Uchida}}
\begin{figure}[t]
 \includegraphics[width=\columnwidth]{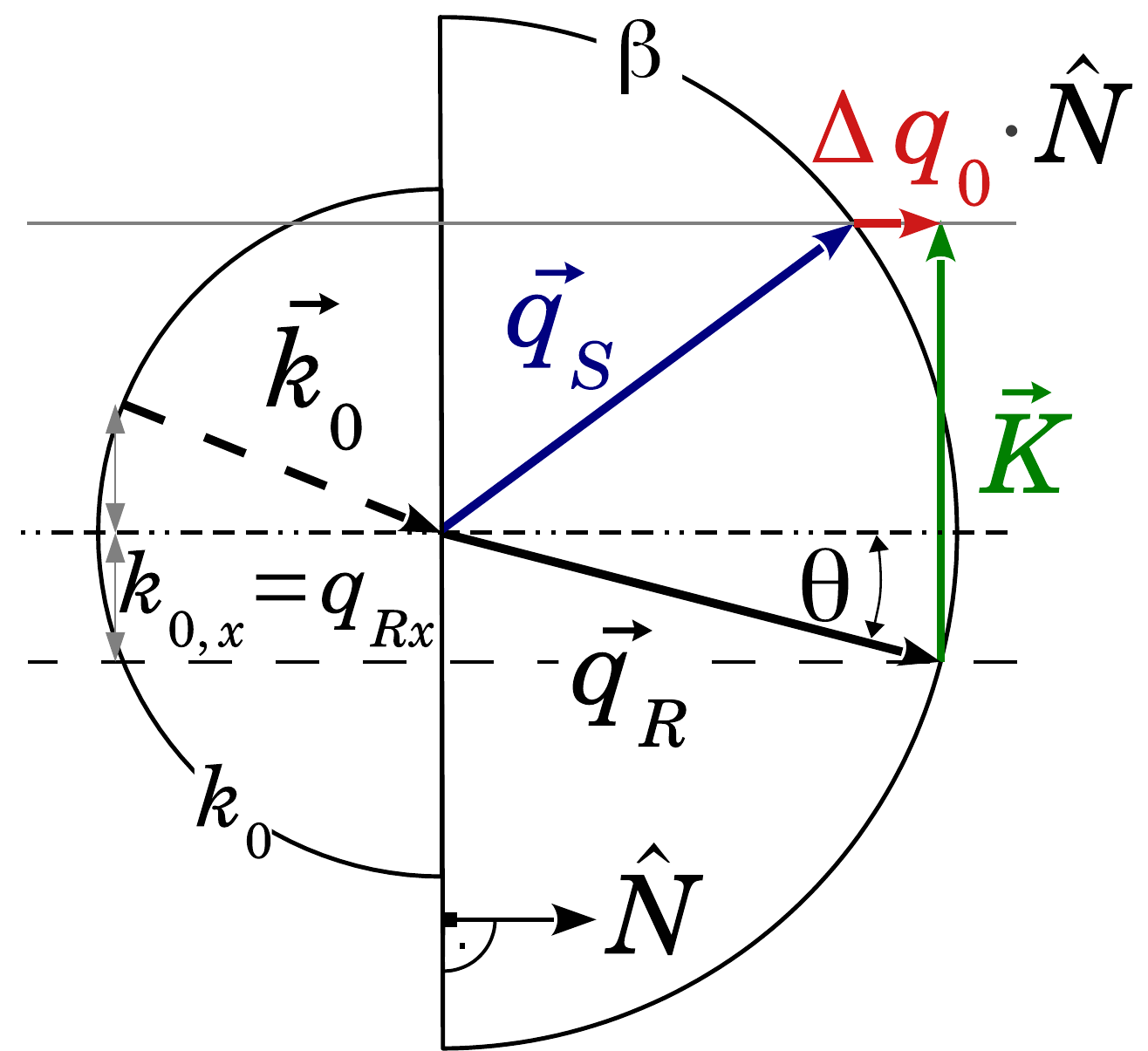}
 \caption{(Color online) Wavevector diagram for the BVM (Uchida) in off-Bragg position with $\beta=|\vec{q}_R|=|\vec{q}_S|$ according to Ref. \cite{Uchida-josa73}.}
 \label{fig:wvUch}
\end{figure}

The only difference between Uchida's and Kogelnik's approach is the choice of the diffracted wave vector as
\begin{equation}\label{eq:kmatchUch}
 \vec{q_S}=\vec{q_R}\pm\vec{K}+\Delta q_0\hat{N}.
\end{equation}
The BVM ensures energy conservation, i.e., 
$|\vec{q}_S|=|\vec{q}_R|=\beta$, by introducing a phase-mismatch parameter $\Delta q_0$. The resulting wavevector diagram for the off-Bragg  case ($\Delta q_0\not=0$) can be seen in Fig. \ref{fig:wvUch}.
From geometrical reasoning the phase mismatch amounts to $\Delta q_0=\beta(c_R-c_S)$ with $c_S=\sqrt{\cos^2\theta+2\vartheta/\beta}=c_R\sqrt{1+2X}$ and a parameter $X=\vartheta/(\beta c_R^2)$, which is convenient for approximations discussed later (Appendix \ref{app:2}).
Solving Eq. (\ref{eq:waveequation}) with the wavevector choice of Eq. (\ref{eq:kmatchUch}) yields for the first order diffraction efficiency
\begin{eqnarray}\label{eq:DEUchida}
\eta_{B}&=&\frac{c_R}{c_S}\left[\nu_{B} {\rm sinc}\sqrt{\nu_{B}^2+\xi_{B}^2}\right]^2\\
\nu_{B}&=&\kappa\frac{d}{\sqrt{c_Rc_S}} \label{eq:nuUchida}\\
\xi_{B}&=&\Delta q_0\frac{d}{2} \label{eq:xiUchida}.
\end{eqnarray}
These equations that have been derived, e.g., in Ref. \cite{Sheridan-jmo92} look quite similar to the ones obtained for Kogelnik's approach. {\em Uchida}'s seminal paper already suggested the BVM \cite{Uchida-josa73}. However, it was somewhat hidden by the second important topic addressed: the attenuation of the grating modulation along the sample depth. The latter is not included in the present treatment in order to enable a direct comparison of the theories. 
\subsection{Two-wave modal approach: Dynamical theory of diffraction (DDT) \label{sec:DDT}}
\begin{figure}
\includegraphics[width=\columnwidth]{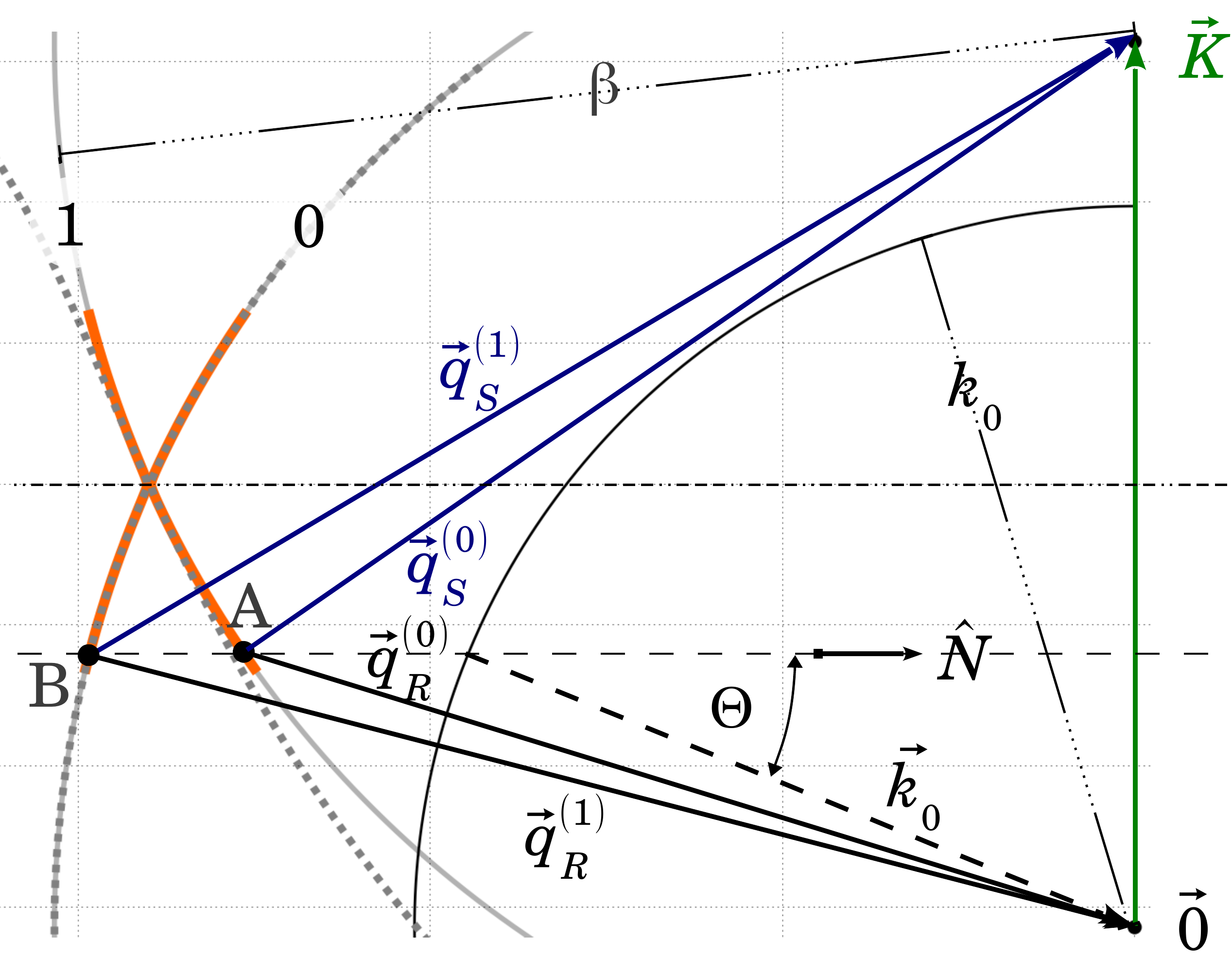}
\includegraphics[width=\columnwidth]{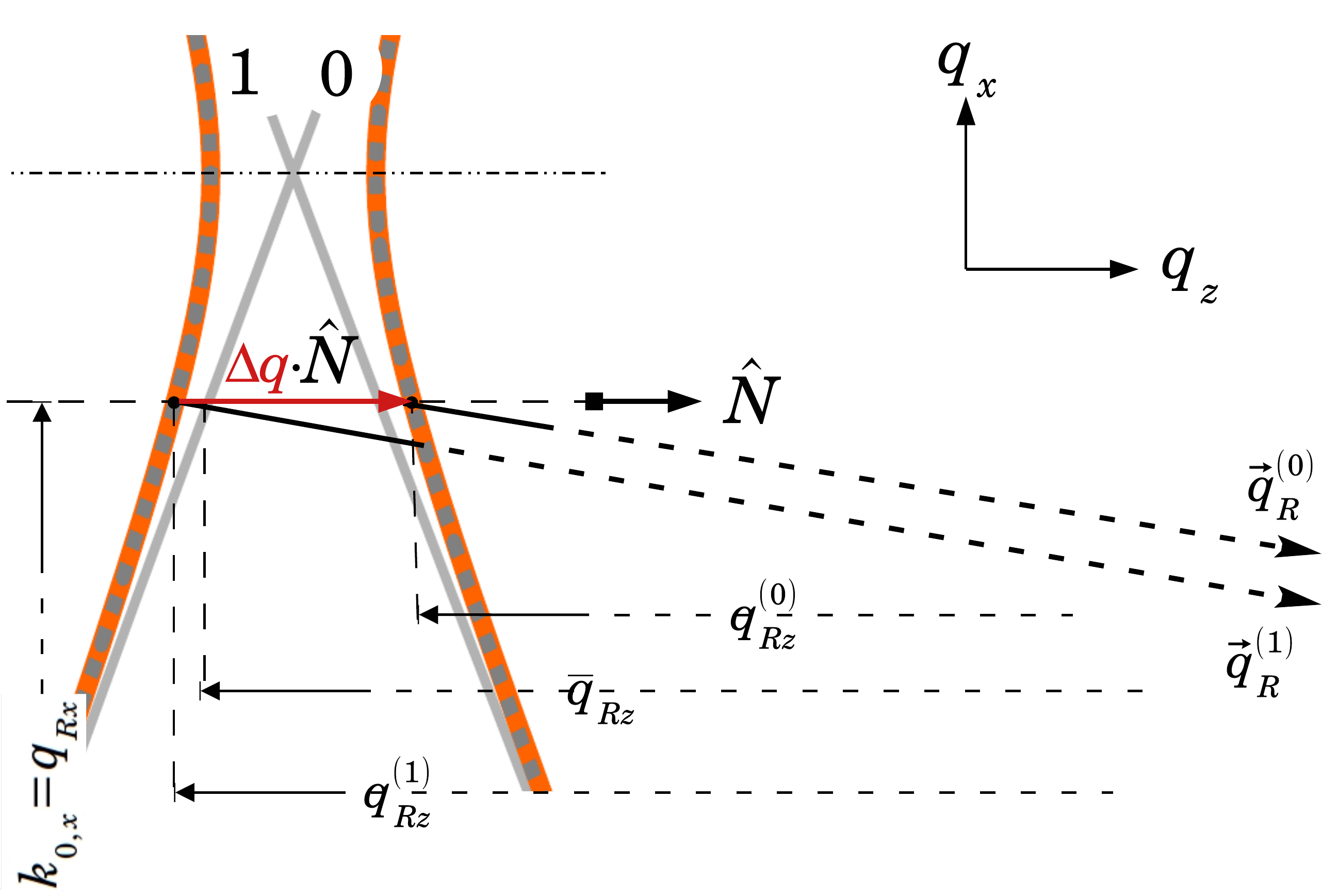}
 \caption{(Color online) Dispersion surface in the off-Bragg geometry. Top: overview including the four permitted wave vectors $\vec{q}_{R;S}^{\,(0);(1)}$. 
 The orange lines represent part of the dispersion surface. An approximate version of the latter - as usually assumed in literature - is shown in dotted gray line style. Bottom: A closeup of the dispersion surface in the vicinity of the Bragg condition with the mismatch $\Delta q$. $\Delta q_0$ is the difference between the gray asymptotic lines (viz spheres) along the $\hat N$-direction, i.e., horizontal.\label{fig:dispsurf}}
\end{figure}

The modal approach to solve Eq. (\ref{eq:waveequation}) is well known in solid state physics for electrons in a periodic potential (band structure, e.g., \cite{Bethe-adp28,Ashcroft-76}), and also for diffraction of X-rays (e.g.,\cite{Ewald-adp16.1,Zachariasen-45,Batterman-rmp64,Pinsker-78}) or neutrons \cite{Rauch-78} by crystal lattices. In volume holography it has been much less prominent \cite{Tamir-cjp66,Sheppard-ije76,Russell-prep81}. The reason might be that it can be usefully applied only for highly idealised gratings \cite{Russell-prep81}. In the modal approach the solution is taken rigorously in the form of a sum of $m$ Bloch waves (eigen-modes) with a grating periodic amplitude function ${\cal E}_{\vec k}$
\begin{eqnarray}
 E^{m}(\vec x)&=&\sum_{\vec k} {\cal E}_{\vec k}^{m} e^{\imath \vec k \vec x},\label{eq:solBloch}\\
 {\cal E}_{\vec k}^m e^{\imath s\vec K}&=&{\cal E}_{\vec k+s \vec K}^m \quad\mbox{with}\quad s\in\mathbb{Z}.\nonumber 
\end{eqnarray}
Fourier transforming Maxwell's equations and inserting Eqs. (\ref{eq:solBloch}) and (\ref{eq:grating}) yields a system of coupled algebraic equations for the amplitudes ${\cal E}_{\vec k}^m$. Details can be found in Appendix \ref{app:1} and Ref. \cite{Russell-prep81}. For sinusoidal volume holograms we know that only four of the amplitudes - two for each eigen-mode - have appreciable magnitude and thus a linear system of two equations for each eigen-mode remains. To obtain consistent and non-trivial solutions to the systems, conditions for the magnitudes of the permitted wavevectors in the grating region arise:
\begin{equation}\label{eq:dispsurf}
\left(2\beta\kappa\right)^2=\left(\beta^2-|\vec{q}_R|^2\right)\left(\beta^2-|\vec{q}_S|^2\right).
\end{equation}
The latter is the decisive equation in the DDT and represents the so-called dispersion surface, i.e., the surface of permitted wavevectors. The dispersion surface and the wave vectors of the eigen-modes in the grating region are shown in Fig. \ref{fig:dispsurf}. These permitted wave vectors can easily be found geometrically by following the three steps below:
\begin{enumerate}
  \item Starting from the origin of the reciprocal space $\vec 0$ draw the incident wavevector with length $k_0$ at the (external) angle of incidence $\Theta$.
  \item Phase matching at the boundary requires that the tangential compontents of the wavevectors are identical, i.e., $k_{0,x}=q_{R,x}$. Thus draw a line along the sample surface normal $\hat N$ and find the points of intersection $A,B$ with the dispersion surface (typically called 'tie-points', two in the transmission geometry)
  \item The permitted wavevectors of propagation are now constructed by forming the vectors $\overrightarrow{AK}=\vec{q}_{S}^{\,(0)},\overrightarrow{BK}=\vec{q}_{S}^{\,(1)},\overrightarrow{A0}=\vec{q}_{R}^{\,(0)},\overrightarrow{B0}=\vec{q}_{R}^{\,(1)}$
\end{enumerate}
The circles with radius $\beta$ around the reciprocal lattice points $\vec{0},\vec{K}$ represent the surface of permitted wavevectors in the medium without a grating viz. $\kappa=0$. They approximate the permitted wavevectors  of propagation in a grating very well unless in the very vicinity of the Bragg condition. This latter situation is shown in the closeup of the dispersion surface in the lower part of Fig. \ref{fig:dispsurf}.

The moduli of the permitted wave vectors of propagation in the grating region are obtained by using Bragg's condition $\vec{q}_S=\vec{q}_R\pm\vec{K}$ and solving Eq. (\ref{eq:dispsurf}) for $\vec q_R$
\begin{equation}\label{eq:permwv}
 \left|\vec{q}^{\,(m)}_R\right|^2=\beta\left(\vartheta+\beta\pm\sqrt{\vartheta^2+(2\kappa)^2}\right).
\end{equation}
This means, that for each propagation direction under consideration ($\vec q=\vec 0,\vec K$) there are two eigen-modes $m=0,1$ with corresponding propagation constants viz. energies or  - as initially pointed out by {\em Ewald} - a sort of birefringence.

Usually, the region in the vicinity of the Bragg condition is approximated, i.e., the spheres around the reciprocal lattice points $\vec 0,\vec K$ with radius $\beta$ are approximated by planes (asymptotic gray lines in Fig. \ref{fig:dispsurf}) and the dispersion surfaces are hyperbolae. We disregard this simplification so that the theory can be applied for the far-off-Bragg region, too. For neutrons the corresponding equations for the diffracted intensities were established in Ref. \cite{Lemmel-prb07,Lemmel-aca13}.
The amplitudes ${\cal E}^m_{\vec k}$ are determined by taking into account the boundary conditions and can be expressed by the grating parameters. Then the first order diffraction efficiency according to the DDT takes the form
\begin{equation}\label{eq:DEDDT}
\eta_{D}=\frac{(2\kappa)^2}{(2\kappa)^2+\vartheta^2}\sin^2\left(\frac{1}{2}\Delta q d\right)
\end{equation}
where $\Delta q=q_{Rz}^{(1)}-q_{Rz}^{(0)}$. This is the function used for the fits to the experimental data in Sec. \ref{sec:Experimental}. 
\section{Experimental \& Results}\label{sec:Experimental}
\begin{figure}
 \centering
 \includegraphics[width=\columnwidth]{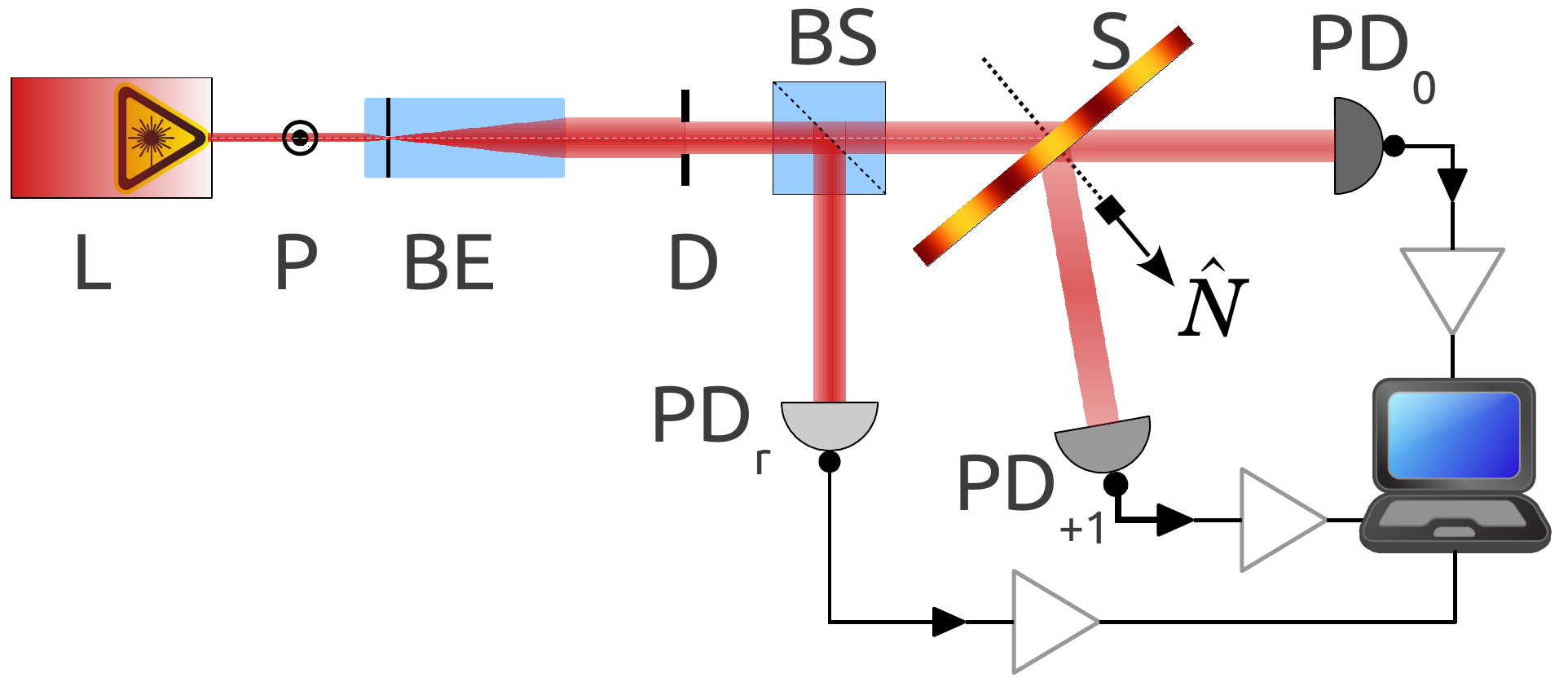}
\includegraphics[height=3.5cm]{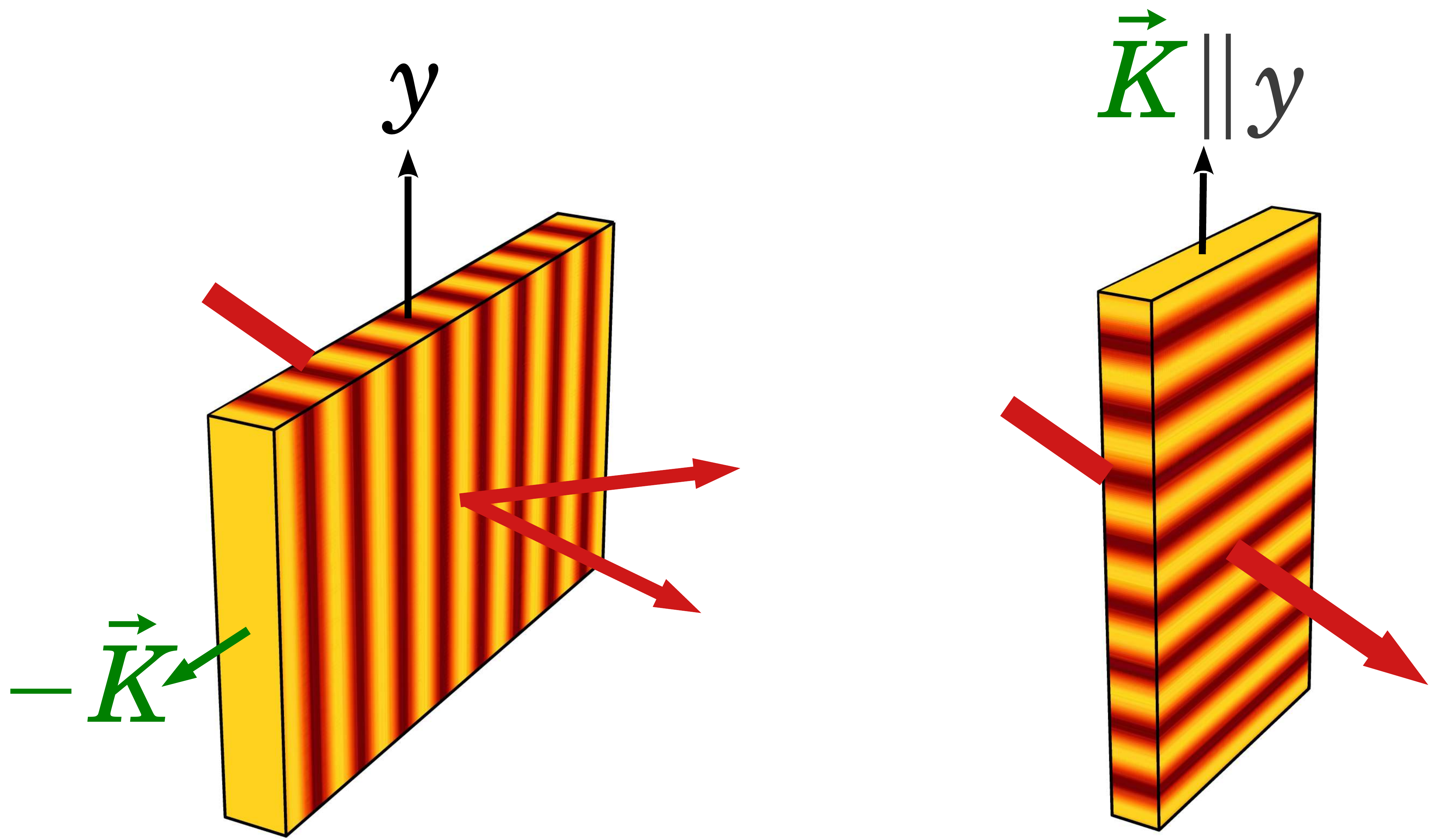}
 \caption{(Color online) Experimental setup scheme: L,P,BE,D,BS,S,PD denote the He-Ne laser, polarizer, beam expander, diaphragm, beamsplitter, sample and photodiodes, respectively. The amplifier is symbolized by the triangle (top). Sample geometry for measuring the diffraction efficiency (left bottom) and the background (right bottom), rotation axis was always $y$.}
 \label{fig:geo}
\end{figure}

\begin{figure}
\includegraphics[width=\columnwidth]{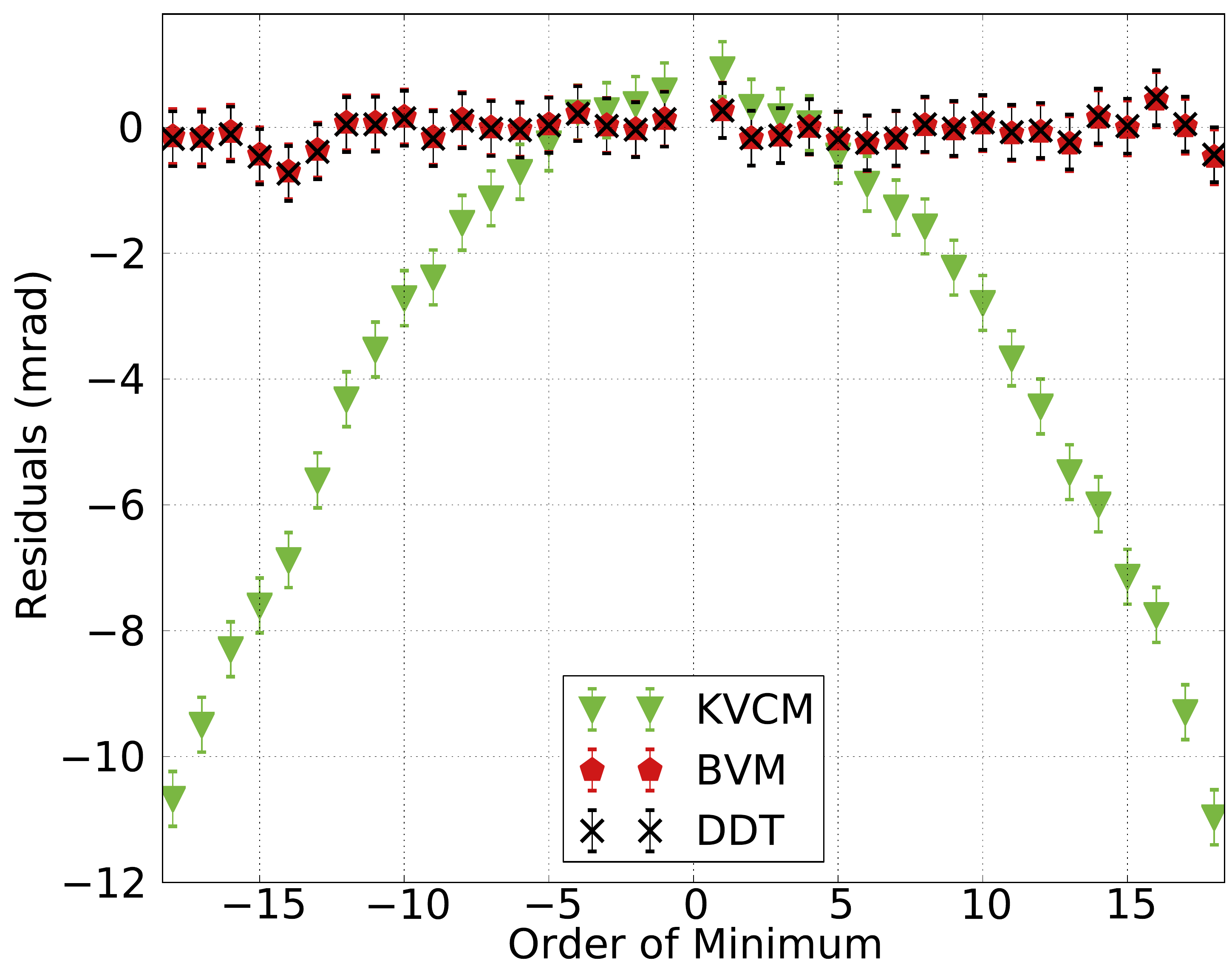}
\caption{(Color online) \label{fig:minima} Comparison of the minimum positions between the experimental data and the three models. 
The deviation of the measured minimum positions from the three fitted models (residuals) is shown.}
\end{figure}

The diffraction experiments were performed on a $\rm SiO_2$ nanoparticle-polymer grating \cite{Suzuki-ao04} with a grating spacing $\Lambda=2\pi/|\vec{K}|=500$ nm using an $s$-polarized He-Ne laser beam at a wavelength $\lambda$ of $632.8$ nm. The sample is the same as used for the experiments concerning the direction of the diffracted beam discussed in Ref. \cite{Fally-apb12}. 
A schematic of the setup is shown in Fig. \ref{fig:geo}. 
The beam was prepared using a standard beam expansion system (microscope objective lens, pinhole, magnification lens) followed by a diaphragm with a diameter of about 5 mm. The sample is placed on a rotation stage (resolution 1/1000 deg) and is rotated around an axis (vertical) parallel to the  $y-$axis while measuring the zero and first order diffracted powers, i.e., rocking curves are recorded. To get reliable values in the far-off-Bragg regime we measured the diffraction powers using Si-photodiodes which were plugged into a light amplifier allowing linear amplification in the $10^{5}$-order signal range. To ensure most reliable results we took at least five power values at each angular position $\theta$ thus being able to evaluate a standard error of the mean. Furthermore, a background measurement was conducted by first rotating the sample by 90$^\circ$ around the sample surface normal, and then recording a rocking curve again, this time the grating vector $\vec{K}$ being parallel to the rotation axis $y$.
 By attaching an opaque circular mask (diameter 5 mm) to the sample's front-surface, it was assured that the incident beam passed the identical sample volume for both, the diffraction experiment and the background measurement.
\begin{figure*}
\includegraphics[width=\columnwidth+\columnwidth]{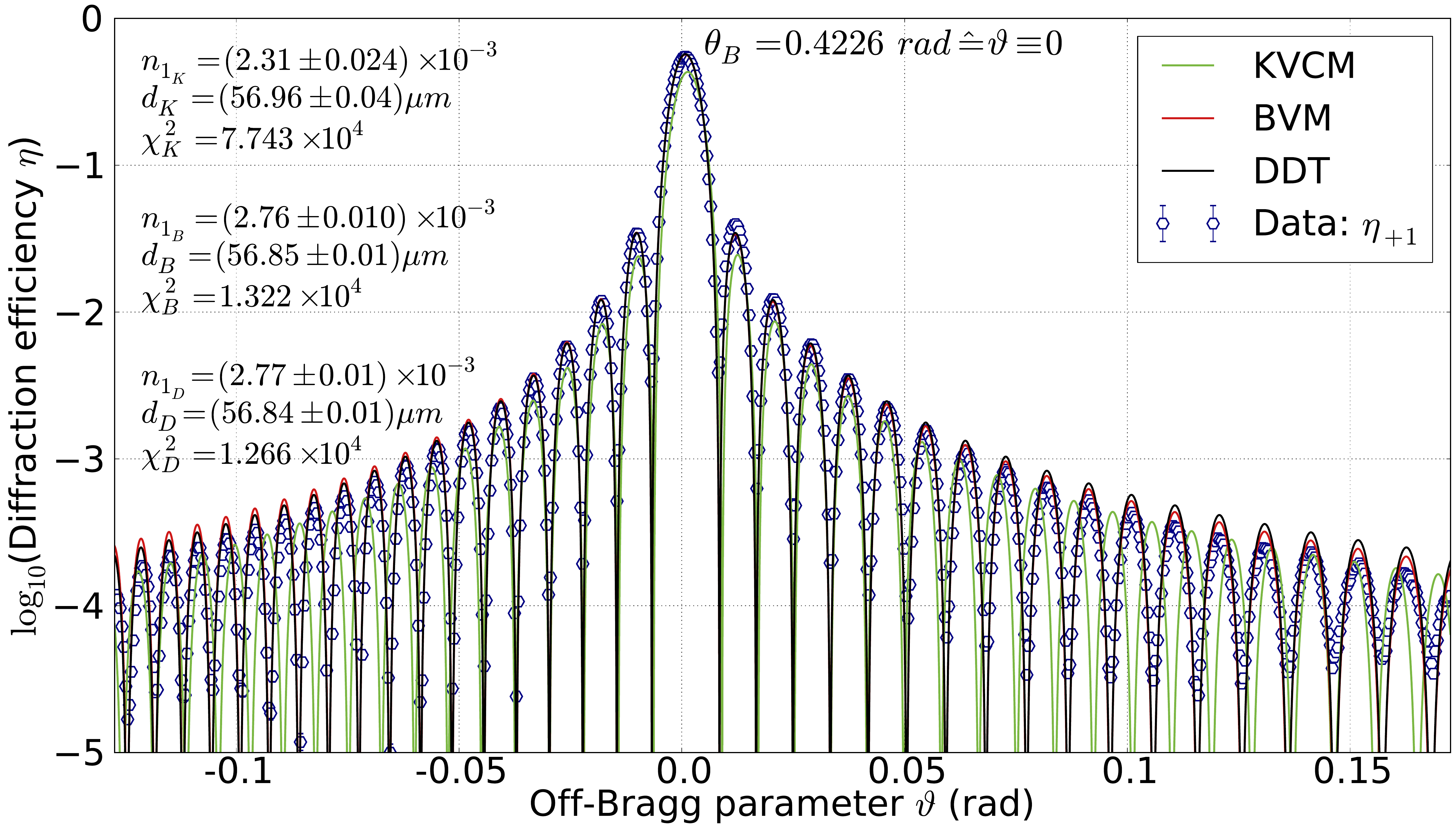}
\caption{(Color online) Dependence of the first order diffraction efficiency on the Off-Bragg parameter $\vartheta$ (see Eq. (\ref{eq:OffBraggparameter})) for a nanoparticle-polymer grating. The plus and minus first order data  (symbols) were fitted using either the KVCM (green line), BVM (red line) or the DDT (black line). Fitting results including the weighted $\chi^2$ are given in the inset. For further details, see text. \label{fig:plusfirst}}
\end{figure*}
The results of the diffraction experiments are shown in Fig. \ref{fig:plusfirst}. A simultaneous least-squares fit to the weighted and background-corrected diffraction efficiency for the minus and plus first order were performed in the angular range $|\theta-\theta_B|\gtrsim 0.15$ radian, therefore including the far-off-Bragg region with 18 side minima. It is obvious from the plot that while the KVCM does not fit the experimental data, BVM and DDT do. This is quantified by the $\chi^2_K$ value being nearly an order of magnitude larger than other approaches.
The most striking difference is observed in the positions of the minima. In Fig. \ref{fig:minima} the experimental minimum positions are compared to those obtained from the theories.
From the angular dependence of the residuals we conclude that the KVCM strongly deviates in the far-off-Bragg region. 
\section{Discussion}
As already said in the introduction, typically the dispersion surface in the DDT is approximated in the vicinity of the Bragg condition so that the dispersion surfaces are hyperbolae. This of course is not applicable for the far-off-Bragg region. On the other hand, a direct comparison of analytical formulae is only possible if proper approximations are applied, which is done in Appendix \ref{app:2}.

Let us start our discussion with mentioning that $\Delta q$ (appearing in DDT) is equivalent to $\Delta q_0$ (appearing in the BVM) in the limit of zero coupling ($\kappa=0$).
In the wavevector diagram (Fig. \ref{fig:wvUch}) a background refractive index $n_0$ is implicitely assumed to describe the permitted wavevectors of propagation in the medium. On the other hand, we know that at the exact Bragg condition a wave experiences a different refractive index within the limits $n_0\pm n_1$. This is only reflected in the DDT, where - in the frame of the two mode case - two waves with slightly different refractive indices propagate towards each of the two reciprocal lattice points $\vec{0}$ and $\vec{K}$ shown in Fig. \ref{fig:dispsurf}. The value of $\Delta q_0$ is the difference between the two crossing asymptotic gray lines (i.e. actually circles) along the sample surface normal shown in the closeup of Fig. \ref{fig:dispsurf}.

In Ref. \cite{Fally-apb12} the effective thickness of the grating - for the same sample - was estimated to be $d_0=58.5\pm0.05 \mu$m by fitting Eq. (\ref{eq:DEUchida})  to the angular dependence of the diffraction efficiency shown in Fig. \ref{fig:plusfirst}. The discrepancy to the value determined in the present work originates from three facts: (i) here we used experimentally measured values weighted by their standard error of the mean (no weighting in \cite{Fally-apb12}), (ii) the angular range was extended to the far-off-Bragg region which has particular influence on the obtained thickness via the phase function through its minima, and (iii) the background was thoroughly subtracted.
Browsing through the literature of evaluating the refractive-index modulation of volume holographic gratings, we find mainly two approaches: a) measuring a rocking curve and calculating $n_1$ and $d$ by using the KVCM in the linearized version and b) just assuming the grating thickness to be the measured mechanical sample thickness and calculating $n_1$ from the diffraction efficiency at the (supposed to be) Bragg angle. While it is evident that the latter can only serve as a rough estimate, the results of the former are usually taken as serious parameters. When fitting of the data shown in Fig. \ref{fig:plusfirst} (without weighting) is performed on this basis, i.e., less thorough, values around the Bragg peak contribute the most to the result. This paradoxically leads to fitting parameters which are much closer to the ones obtained by BVM and DDT (deviations of about 1\% for $n_1$ and $d$). The latter, however, is only due to the fact that we have an almost ideal grating very accurately described by Eq. (\
ref{eq:grating}).

After all we are left with the question: are we able to decide if any of the theories is superior? Our experiment and the sample under investigation provide almost ideal conditions, i.e., the assumptions about the grating taken in the models are fulfilled. We identify the KVCM as less reliable
due to the significantly higher $\chi^2_K\gg\chi^2_{{B},D}$ (see Fig. \ref{fig:plusfirst}). A decision on the basis of $\chi^2$ alone to recommend either the BVM or the DDT is not evident. 
However, when it comes to less perfect gratings, e.g., taper with an attenuation of the grating modulation along the sample depth \cite{Suzuki-ao07} due, for example, to non-neglible linear absorption during recording \cite{Tomita-apl03}, there is no known way to apply the modal approach that leads to the DDT. In contrast, it has been successfully proven that the problem can even be analytically treated using a coupled wave analysis \cite{Uchida-josa73}. 
\section{Conclusion}
Diffraction by a one dimensional volume holographic grating with a relatively thick film (about 57 micron) and high refractive- index modulation ($n_1$ about $0.003$) allows to demonstrate a significant difference between diffraction theories. 
For now we conclude, that both coupled wave approaches, KVCM and BVM, and the corresponding differential equations are  approximations to the DDT. While the KVCM fails to give the correct phase function for the diffraction efficiency, the BVM differs by a factor $c_R/c_S$ in the amplitude. We finally comment that from a practical point of view most of the gratings are not that perfect, and thus the BVM including the attenuation along sample depth as discussed in Ref. \cite{Uchida-josa73} should be the first choice.
\section*{Acknowledgments}
This work was supported by the Austrian Science Fund (P-20265), the Slovenian-Austrian bilateral programme \"OAD-WTZ  (SI 07/2011) and by  the Ministry of Education, Culture, Sports,
Science, and Technology of Japan (Grant No. 23656045). 
\appendix
\section{Derivation of the dispersion surface Eq. (\ref{eq:dispsurf})\label{app:1}}
We follow the derivations given by {\em Batterman and Cole} for x-rays \cite{Batterman-rmp64} and {\em Russell} for light \cite{Russell-prep81}.

We start with the Helmholtz equation (\ref{eq:waveequation}). The refractive index from Eq. (\ref{eq:grating}) is a real quantity and periodic with the reciprocal lattice vector $\vec{K}=(K,0,0)$; the relative permittivity of the medium is defined as usual, i.e., $n^2=\varepsilon\in\mathbb{R}$. Hence,
\begin{equation}\label{eq:vareps}
  \varepsilon(\vec x)=\varepsilon_0+\frac{1}{2}\sum_{g\in\mathbb{N},g>0}\varepsilon_g\left(e^{\imath \vec K_g\vec x}+e^{-\imath \vec K_g\vec x}\right),
\end{equation}
where $\vec K_g\equiv g \vec K, g\in\mathbb{N}$. Further, it is known that solutions to the wave equation for periodic media have a particular form 
\begin{equation}\label{eq:Blochwave}
  E^m(\vec x)=\sum_{\vec k}{\cal E}_{\vec k}^m e^{\imath\vec k \vec x}
\end{equation}
where ${\cal E}_{\vec k}^m$ are functions with the periodicity of the medium. The $E^m(\vec x)$ are eigen-modes of the wave equation labelled by $m$ (indicating different energies numbered in increasing order) and are called Bloch waves. By inserting Eqs. (\ref{eq:vareps}) and (\ref{eq:Blochwave}) into Eq. (\ref{eq:waveequation}) we end up with an infinite number of algebraic equations for the infinite number of coefficients ${\cal E}_{\vec k}^m$:
\begin{widetext}
\begin{eqnarray}
\sum_{\vec k}\left[(\varepsilon_0k_0^2-|\vec k|^2){\cal E}_{\vec k}^m e^{\imath\vec k \vec x}+\frac{k_0^2}{2}
\sum_{g\in\mathbb{N},g>0}\varepsilon_g\left(e^{\imath (\vec k+\vec K_g)\vec x}+e^{\imath (\vec k-\vec K_g)\vec x}
\right){\cal E}_{\vec k}^m
\right]&=&0\nonumber\\
\sum_{\vec k}\left[(\beta^2-|\vec k|^2){\cal E}_{\vec k}^m+\frac{k_0^2}{2}
\sum_{g\in\mathbb{N},g>0}\varepsilon_g({\cal E}_{\vec k+\vec K_g}^m+{\cal E}_{\vec k-\vec K_g}^m)\right]&=&0.
\end{eqnarray}
\end{widetext}
This is effectively the Fourier transformed version of the wave equation in a periodic medium.

Assuming that only two of the coefficients considerably differ from zero - say ${\cal E}_{\vec q}^m,{\cal E}_{\vec q+\vec K}^m$ - the system of equations reduces to
\begin{eqnarray}\nonumber
(\beta^2-|\vec q|^2) {\cal E}_{\vec q}^m+\frac{k_0^2}{2}\varepsilon_1 {\cal E}_{\vec q+\vec K}^m&=&0\quad\mbox{for}\quad\vec k=\vec q\\
\frac{k_0^2}{2}\varepsilon_1 {\cal E}_{\vec q}^m+(\beta^2-|\vec q+\vec K|^2) {\cal E}_{\vec q+\vec K}^m&=&0\quad\mbox{for}\quad\vec k=\vec q+\vec K\nonumber
\end{eqnarray}
Nontrivial solutions to this system of equations require the determinant of the coefficient-matrix ${\cal E}_{\vec k}^m$, i.e., the characteristic polynomial, to vanish. This leads to a quartic equation in $|\vec q|$
\begin{equation}\label{eq:biquadrat}
  (\beta^2-|\vec q|^2)(\beta^2-|\vec q+\vec K|^2)=\left(\frac{k_0^2}{2}\varepsilon_1\right)^2,
\end{equation}
which gives the moduli $q^{(m)}$ of the $m$ permitted wavevectors for propagation in a periodic medium. The latter is Eq. (\ref{eq:dispsurf}) when using the relations $\kappa=k_0n_1/2, \varepsilon_1=2n_1n_0$ and the notation $\vec q=\vec q_R, \vec q+\vec K=\vec q_S$.

Finally, we find the total field in the grating as a sum over all eigen-modes
\begin{equation}
\label{eq:totalfield}
E(\vec x)=\sum_m v_m\sum_{\vec k=\vec q^{(m)},\vec q^{(m)}+\vec K} {\cal E}_{\vec k}^m e^{\imath \vec k \vec x}, 
\end{equation}
where the coefficients $v_m$ must be determined considering the boundary conditions. We express the ratio
\begin{equation}
\frac{{\cal E}_{\vec q}^m}{{\cal E}_{\vec q+\vec K}^m}=\frac{2\kappa}{\vartheta\pm\sqrt{\vartheta^2+(2\kappa)^2}},  
\end{equation}
and assume that the incident wave at the entrance boundary has an amplitude equal to unity so that Eq. (\ref{eq:totalfield}) yields conditional equations for the $v_m$:
\begin{eqnarray}
  v_0&=&\frac{{\cal E}_{\vec q+\vec K}^1}{{\cal E}_{\vec q+\vec K}^1{\cal E}_{\vec q}^0-{\cal E}_{\vec q}^1{\cal E}_{\vec q+\vec K}^0} \nonumber\\
  v_1&=& -v_0\frac{{\cal E}_{\vec q+\vec K}^0}{{\cal E}_{\vec q+\vec K}^1}\nonumber.
\end{eqnarray}
The field-amplitude of the waves travelling towards the reciprocal lattice point $\vec K$ is
\begin{equation}
E_{-1}(\vec x)=v_0{\cal E}_{\vec q+\vec K}^0 e^{\imath (\vec q^{(0)}+\vec K)\vec x}+v_1{\cal E}_{\vec q+\vec K}^1e^{\imath (\vec q^{(1)}+\vec K)\vec x}
\end{equation}
so that for the geometry used in the experiment discussed here (Fig. \ref{fig:geo}) the diffracted amplitude $E_{-1}(z=d)$ amounts to
\begin{eqnarray}\nonumber
  E_{-1}&=&\frac{(|\vec q^{\,(0)}|^2-\beta^2)(|\vec q^{\,(1)}|^2-\beta^2)}{2\beta\kappa(|\vec q^{\,(1)}|^2-|\vec q^{\,(0)}|^2)} \\
  &\times& \left(e^{\imath q_{z}^{(0)}d}-e^{\imath q_{z}^{(1)}d}
  \right).
\end{eqnarray}
Using the definition $\eta=|E_{-1}(z=d)/1|^2$ we end up with Eq. (\ref{eq:DEDDT}).
\section{'Proper' approximation in the DDT for comparison with KVCM and BVM\label{app:2}}
To obtain an equation similar to the form of Eqs. (\ref{eq:DEKogelnik}) and (\ref{eq:DEUchida}) also for the DDT, we approximate $\Delta q$ using Eq. (\ref{eq:permwv}):
\begin{eqnarray}\nonumber
q_{Rz}^{(m)}-\overline{q}_{Rz}&\approx&\sqrt{\beta(\vartheta+\beta\cos^2\theta)}\left[1\pm\frac{\sqrt{\vartheta^2+(2\kappa)^2}}{2(\vartheta+\beta\cos^2\theta)}\right],\nonumber\\
\label{eq:permapp1}
 \Delta q&\approx&\sqrt{\beta}\sqrt{\frac{\vartheta^2+(2\kappa)^2}{\vartheta+\beta c_R^2}}.
\end{eqnarray}

Then the functional form of the diffraction efficiency - an amplitude term $A$ multiplied by an oscillatory phase term $\sin^2\Phi$ - looks similar to that of Eqs. (\ref{eq:DEKogelnik}) and (\ref{eq:DEUchida}) yielding
\begin{eqnarray}\label{eq:DEDDT1}
\eta_{D}&=&A_D \sin^2{\Phi_D}\\
A_D&=&\frac{(2\kappa)^2}{\vartheta^2+(2\kappa)^2}\nonumber\\
&=&\frac{(2\kappa)^2}{(\beta c_R^2 X)^2+(2\kappa)^2}
=\frac{\nu_K^2}{\xi_K^2+\nu_K^2}\label{eq:amplitudeD}\\
\Phi_D&=&\sqrt{\frac{1}{1+X}}\sqrt{(\beta c_R^2 X)^2+(2\kappa)^2}\frac{d}{2c_R}\nonumber\\
&=&\sqrt{\frac{1}{1+X}}\sqrt{\nu_K^2+\xi_K^2}.\label{eq:PhiD}
\end{eqnarray}
For the experimental angular range $\theta_B$-0.155\ldots$\theta_B$+0.166 radian (angles in the medium) and $\kappa=2.08\times 10^{-2}\mu{\rm m}^{-1}$ (as obtained from the fit) the deviation of Eq. (\ref{eq:DEDDT1}) from Eq. (\ref{eq:DEDDT}) is less than about 0.3\%. Rewriting  Eq. (\ref{eq:DEKogelnik}) for comparison in terms of $X,\nu_K,\xi_K$, we have
\begin{eqnarray}\label{eq:DEKogelnik1}
 \eta_K&=&A_K\sin^2{\Phi_K}\\
A_K&=&A_D\label{eq:amplitudeK}\\
\Phi_K&=&\Phi_D\sqrt{1+X}.\label{eq:PhiK}
\end{eqnarray}
Thus, the phase term of the KVCM differs from that of the DDT by a factor $\sqrt{1+X}$ while the amplitudes are identical. We proceed with Eq. (\ref{eq:DEUchida}) in terms of $X$ which yields
\begin{eqnarray}
\eta_{B}&=&A_{B}\sin^2{\Phi_{B}}\\ \nonumber
A_{B}&=&\frac{1}{\sqrt{1+2X}}\\
&\times&\frac{(2\kappa)^2}{(2\kappa)^2+(\beta c_R^2(1-\sqrt{1+2X}))^2\sqrt{1+2X}}\label{eq:amplitudeU}\\\nonumber
\Phi_{B}&=&\frac{d}{2c_R}\nonumber\\
&\times&\sqrt{[\beta c_R^2(1-\sqrt{1+2X})]^2+\frac{(2\kappa)^2}{\sqrt{1+2X}}}.\label{eq:PhiU}
\label{eq:DEUchida1}
\end{eqnarray}
Despite the formal similarity of the expressions Eqs. (\ref{eq:DEKogelnik}) and (\ref{eq:DEUchida}) for KVCM and BVM, respectively, we 
need to take a further approximation for a direct comparison. We recall that we are interested in the far-off-Bragg region, where the theories might significantly differ one another. Then $\vartheta\gg(2\kappa)$ and we neglect $\kappa$ in Eqs. (\ref{eq:DEDDT1}) - (\ref{eq:DEUchida1}) in comparison to $\vartheta$ and $X$. The approximate phase functions then read
\begin{eqnarray}\label{eq:dephasD}
\Phi_{0,D}&=&|X|\sqrt{\frac{1}{1+X}} \frac{\beta c_R d}{2}\\
\Phi_{0,{B}}&=& |1-\sqrt{1+2 X}| \frac{\beta c_R d}{2}\label{eq:dephasB}\\
\Phi_{0,K}&=& |\vartheta| \frac{d}{2c_R}=|X| \frac{\beta c_R d}{2}.\label{eq:dephasK}
\end{eqnarray}
\begin{figure}[t]
 \centering
 \includegraphics[width=\columnwidth]{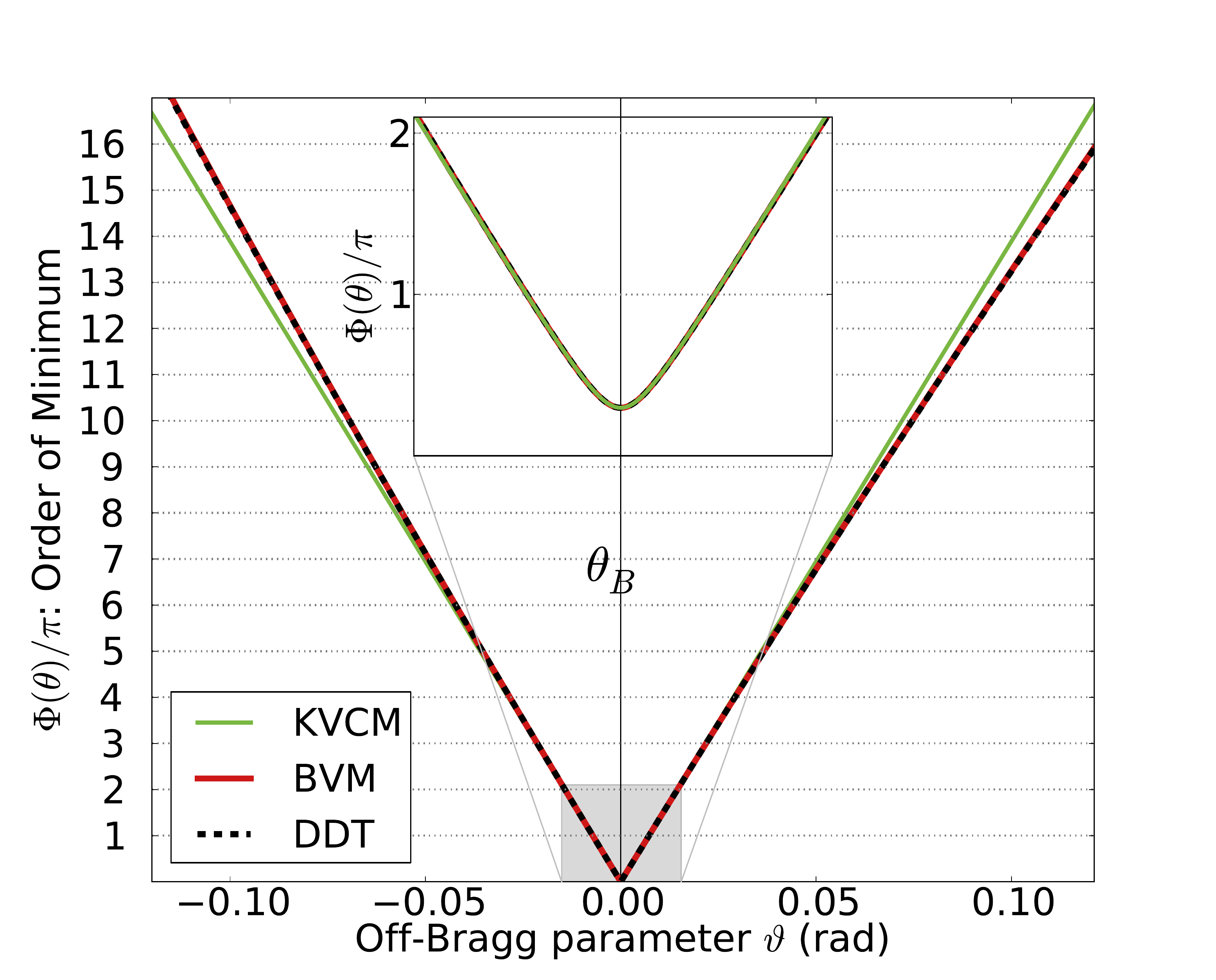}
 \caption{(Color online) \label{fig:phase}$\Phi(\theta)$ for each approach according to Eqs. (\ref{eq:PhiD}),(\ref{eq:PhiK}),(\ref{eq:PhiU}). Roots of the diffraction efficiency occur at angles for which $\Phi(\theta_s)=s\pi$, i.e, where the dotted horizontal lines intersect the functions.}
\end{figure}
Higher order minimum positions are given by $\Phi(\theta_s)=s\pi$ with $s\in\mathbb{N}$. Fig. \ref{fig:phase} shows $\Phi(\theta)$ for each of the theories.
BVM and DDT lead to identical results in the phase function up to second order in $\theta-\theta_B$, whereas KVCM is equivalent only in the very vicinity of the Bragg condition.
For the amplitude part of the diffraction efficiency, the situation is different. Assuming again $\vartheta\gg2\kappa$ (for the far-off-Bragg angular range) we get
\begin{eqnarray}\label{eq:amplitudes}
A_{0,D}&=&\left(\frac{2\kappa}{\beta c_R^2 X}\right)^2\\\nonumber
A_{0,{B}}&=& \frac{1}{\sqrt{1+2X}}\frac{(2\kappa)^2}{(\beta c_R^2 [1-\sqrt{1+2X}])^2\sqrt{1+2X}}\\
A_{0,K}&=&A_{0,D}.\nonumber
\end{eqnarray}
\begin{figure}[t]
 \centering
 \includegraphics[width=\columnwidth]{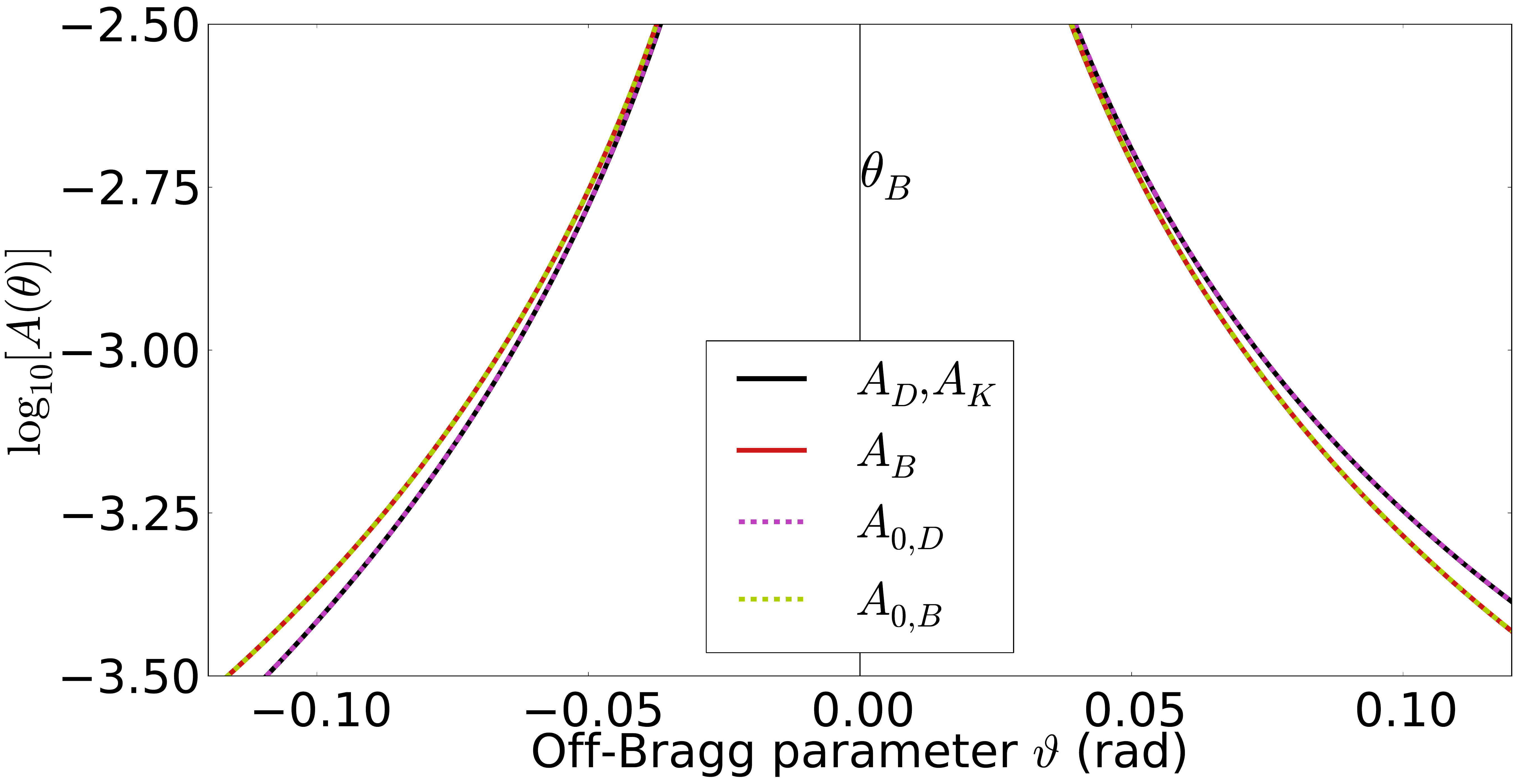}
 \caption{(Color online) The amplitude functions $A(\theta)$ - Eqs. (\ref{eq:amplitudeD}),(\ref{eq:amplitudeK}),(\ref{eq:amplitudeU}) - and $A_0(\theta)$ - Eqs. (\ref{eq:amplitudes})  - in the off-Bragg region for each model.}
 \label{fig:amplitude}
\end{figure}
In Fig. \ref{fig:amplitude} the amplitude functions $A(\theta)$ are shown. This time the DDT is identical to the KVCM and differs from BVM, namely by a factor $\sqrt{1+2X}$.
The relative deviation of $1-\sqrt{1+2X}A_{B}/A_{D}$ is less than $0.7$\% in the range discussed.

To summarize: For the diffraction efficiency
\begin{itemize}
\item the phase term of the DDT agrees to that of the BVM excellently; the KVCM differs by a factor of $(1+X)^{-1/2}$, while
\item the amplitude term of the DDT is equal to that of KVCM; whereas now that of the BVM differs by a factor of $(1+2X)^{1/2}$.
\end{itemize}
While the first statement is fully confirmed by the experimental data and can be seen in Figs. \ref{fig:plusfirst} and  \ref{fig:minima}, the second statement can hardly be verified from the dataset. One reason could be that the side wings of the amplitude function are also influenced by absorption, an issue not considered here.
    \subsection*{Approximate theories and their interrelation\label{app:3}}
As said above, the equation for the dispersion surface Eq. (\ref{eq:dispsurf}) discussed in nearly any publication is approximated to form hyperbolic sheets in the vicinity of the Bragg incidence, i.e., the quartic equation in $\beta$ is transformed in a quadratic one with the approximation $\beta^2-(\vec{q})^2\approx(\beta-|\vec{q}|)2\beta$. Then the magnitudes of the permitted wavevectors are given by
\begin{equation}
\left|\vec{q}^{\,(m)}_R\right|=\beta+\frac{1}{2}\left(\vartheta\pm\sqrt{\vartheta^2+(2\kappa)^2}\right).
\end{equation}
The approximate version of the dispersion surface is shown in Fig. \ref{fig:dispsurf} as dotted grey lines. Such an approximation leads to the identical diffraction efficiency for the DDT and the KVCM as given by Eq. (\ref{eq:DEKogelnik}) \cite{Russell-prep81,Klepp-nima11}. Using the approximation of Eq. (\ref{eq:permapp1}) and truncating the expansion of the phase correction factor $(1+X)^{-1/2}\approx 1+O[X]$ in Eq. (\ref{eq:DEDDT1}) with the constant term, we arrive at the same result. 

In addition, the off-Bragg parameter is frequently linearized as in Ref. \cite{Kogelnik-Bell69}, so that $\vartheta(\theta)=K\cos\theta_B(\theta-\theta_B)$ which then should give even worse results and anyhow is valid only close to the Bragg angle $\theta_B$.
  \bibliographystyle{apsrev}

\end{document}